\documentclass[aps,prd,onecolumn,showpacs,superscriptaddress]{revtex4}
\usepackage{amsmath}
\usepackage{graphicx}

\newcommand{\bastar}{\begin{eqnarray*}}
\newcommand{\eastar}{\end{eqnarray*}}
\newskip\humongous \humongous=0pt plus 1000pt minus 1000pt

\newif\ifdtup

\relax
\newcommand{\be}{\begin{equation}}
\newcommand{\ee}{\end{equation}}
\newcommand{\bea}{\begin{eqnarray}}
\newcommand{\eea}{\end{eqnarray}}

\newcommand{\nn}{\nonumber}

\begin{document}
\title{Bargmann-Wigner Formulation and 
Superradiance Problem of Bosons and Fermions 
in Kerr Space-time}

\author{Masakatsu Kenmoku}
\address{Nara Science Academy, 178-2 Takabatake, Nara 630-8301, Japan  \\
E-mail: kenmoku@asuka.phys.nara-wu.ac.jp}
\author{Y. M. Cho}
\address{Administration Building 310-4,
Konkuk University, Seoul 143-701, Korea}
\address{School of Physics and Astronomy,
Seoul National University, Seoul 151-747, Korea\\
E-mail: ymcho7@konkuk.ac.kr}

\begin{abstract}
The superradiance phenomena of massive bosons and fermions 
in the Kerr spacetime are studied in the Bargmann-Wigner 
formulation. In case of bi-spinor, the four independent 
components spinors correspond to the four bosonic freedom: 
one scalar and three vectors uniquely. The consistent 
description of the Bargmann-Wigner equations between fermions 
and bosons shows that the superradiance of the type with 
positive energy $(0<\omega)$ and negative momentum near 
horizon $(p_{\rm H}<0)$ is shown not to occur. On the 
other hand, the superradiance of the type with negative 
energy $(\omega<0)$ and positive momentum near horizon 
$(0<p_{\rm H})$ is still possible for both scalar bosons 
and spinor fermions.
\end{abstract}


\maketitle



\section{Introduction}

One of standard radiation problems of matter fields around 
rotating black holes is the superradiance problem in which 
the reflected intensity becomes stronger than the incident 
intensity. The successive occurrence of the superradiance 
could cause a serious problem on the stability of black holes 
known as the blackhole bomb \cite{Press72,Chandra83,Cardo04,Koda08}. 
The superradiance phenomena have still new interests in theoretical 
and numerical analysis \cite{East13}. 

Under Kerr space-time background, Klein-Gordon, Dirac, Maxwell, 
Rarita-Shwinger and Einstein equations for massless fields are 
known to reduce to the separable one component field equations, 
which are called as the Teukolsky equations \cite{Teukol72,Teukol74}. 
The analytic perturbative solutions of the Teukolsky equations 
show that the reflected intensity can become stronger than the 
incident intensity for bosons, which tells thatthe superradiance 
occurs for strongly rotating (or light mass) black holes. 

We list up some special features of the superradiance phenomena 
to rotating black holes as follows:  \\ 
1. The superradiance may occur for bosonic fields to the strongly 
rotating (or light mass) black holes but not to the weakly 
rotating (or heavy mass) black holes in four dimensional 
space-time \cite{Press72,Misner72,Unruh73,Unruh74,Det80,Taka97,Muko00}. \\
2. The superradiance does not occur for massive as well as 
massless fermionic fields to any rotating black holes in 
four dimensional space-time \cite{Unruh73,Unruh74,Maeda76,Wagh85}. \\
3. In the three dimensional case, the superradiance phenomena 
of the type with positive energy $(0<\omega)$ and negative 
momentum near horizon $(p_{\rm H}<0)$ have been shown not 
to occur \cite{Birming01,Kenmoku08,Kenmoku08-2}. 

Although this difference between bosons and fermions has 
been attributed to the exclusion principle, we need a 
better understanding. The purpose of this paper is to resolve 
this problem using the Bargmann-Wigner (BW) formulation 
which connects massive fermions and bosons consistently 
in the unquantized field theory. Bosons can be considered 
as composite particles of even number of fermions, and 
the BW formulation realizes the consistent relation among 
them. 

We extend the original BW equations to include scalars 
(lower spin state) as well as vectors (highest spin state) 
in the bi-spinor case. It should be noted that the boson 
states can be described by spinor from one side and by 
boson from the other side in BW formulation. Then we apply 
it to the current conservation law among massive scalars 
and spinors in asymptotic region of Kerr space-time.  
It is noted that the BW equations relate, in this case, 
bi-spinor with bosons kinematically contrasting with 
the Bethe-Salpeter equations which relate them dynamically. 

Our result shows that the momentum near horizon cannot be negative, 
therefore the superradiance for bosons does not occur in case of 
$0<\omega$ and $p_{\rm H}=\omega-m\Omega_{\rm H}<0$ (where $m$ is 
the magnetic quantum number of fields and $\Omega_{\rm H}$ is the 
angular velocity of black hole). This tells that there is no 
difference in superradiance between fermions and bosons. This 
article is partially based on the preliminary work \cite{Kenmoku12}.

The organization of this paper is as follows. In section 2, 
BW equations for bi-spinor in flat space-time are extended 
to include scalar part as well as vector parts. The Lagrangian 
and the conserved currents are also studied. In section 3, 
BW formulation is applied to the asymptotically flat space-time 
and try to solve the fermion and boson puzzle in superradiance 
phenomena in Kerr geometry. The summary is given in the final 
section. 

\section{Bargmann-Wigner formulation in flat spacetime}

From the representation theory of Lorentz group, the general 
spin states can be formulated by the direct product of the 
fundamental representation (spin 1/2 spinors). This is realized 
in the relativistic field equations for arbitrary spin \cite{Dirac36}. 
According to the original formulation of Bargmann and Wigner 
\cite{Barg48,Lurie}, bosonic fields of mass $\mu$ and 
spin s are represented by a completely symmetric multi-spinor 
of rank 2s 
\[
{\Psi(x)}^{(\rm BW)}_{\alpha \beta \dots \tau}
={\Psi(x)}^{(\rm BW)}_{\beta \alpha \dots \tau}
\] 
satisfying Dirac-type equations in all index: 
\begin{eqnarray}
(\gamma^{\lambda} {\partial}_{\lambda}+\mu)_{\alpha \alpha^{'}}
{\Psi(x)}^{(\rm BW)}_{\alpha^{'} \beta \dots \tau}&=&0  \nonumber  \\ 
(\gamma^{\lambda}{\partial}_{\lambda}+\mu)_{\beta \beta^{'}}
{\Psi(x)}^{(\rm BW)}_{\alpha \beta^{'} \dots \tau}&=&0 \nonumber \\ 
\cdots \ . \label{BWequation}
\end{eqnarray}
In the following we do not require the completely symmetry 
for the multi-spinor index of rank 2s in order to include 
any bosonic spin states 0, 1, $\dots$ s. 

Let us start with the Bargmann-Wigner equations for scalar and 
vector fields. For spin 0 and 1 states, we introduce the 
modified second rank BW field by
\begin{eqnarray}
\Psi(x)=\Psi(x)^{(\rm BW)}C^{-1}\gamma_{5}, 
\end{eqnarray}
where $C$ denotes the charge conjugation operation. With this 
the BW equations (\ref{BWequation}) are expressed as
\begin{eqnarray}
(\gamma^{\lambda}\partial_{\lambda}+\mu)\Psi(x)&=&0,  
\label{BWeq1}\\ 
\Psi(x)(\overleftarrow{\partial}_{\lambda}\gamma^{\lambda}
+\mu)&=&0. 
\label{BWeq2}
\end{eqnarray}
We expand the bi-spinor fields with a set of bosons as
\begin{gather}
\Psi(x)=\sqrt{\mu}(S(x)+\gamma_{5}P(x)
-\gamma^{\lambda}V_{\lambda}(x)
+\gamma_{5}\gamma^{\lambda}A_{\lambda}(x)) \nn\\
+\frac{1}{2\sqrt{\mu}}\gamma_{5}
\Sigma^{\lambda\tau}F_{\lambda\tau}(x), 
\label{Bexp}
\end{gather} 
where $S$ and $A_{\lambda}$ denote scalar and vector 
fields respectively, and the spin matrix is defined by 
$\Sigma_{\lambda\tau}:=(\gamma_{\lambda}\gamma_{\tau}
-\gamma_{\tau}\gamma_{\lambda})/2$. Other fields 
$P, V_{\lambda}, F_{\lambda\tau}$ are auxilualy fields. 

Now we apply BW equations (\ref{BWeq1})-(\ref{BWeq2}) to
boson expansion form (\ref{Bexp}). Adding these equations, 
we find the set of relations among boson fields 
\begin{eqnarray}
\mu S(x)&=&\partial^{\lambda}V_{\lambda}(x)\ 
\label{SVrelation} \label{SV}\\
\mu V_{\lambda}(x)&=&\partial_{\lambda}S(x) 
\label{VSrelation}\label{VS}\\
P(x)&=&0 \label{P}\\
\mu^2A_{\lambda}(x)&=&\partial^{\tau}F_{\tau\lambda}(x) 
\label{AF}\\
F_{\lambda\tau}(x)&=&\partial_{\lambda}A_{\tau}(x)
-\partial_{\tau}A_{\lambda}(x). 
\label{FA}
\end{eqnarray}
With this we obtain the Klein-Gordon type field equations
for independent spin 0 and 1 fields,  
\begin{eqnarray}
(\partial^{\lambda}\partial_{\lambda}-{\mu}^2)S(x)&=&0,
\label{Seq1}\\
\partial^{\lambda}(\partial_{\lambda}A_{\tau}
-\partial_{\tau}A_{\lambda})-{\mu}^2A_{\tau}(x)&=&0. 
\label{Aeq1}
\end{eqnarray}
The supplementary condition can be derived from (\ref{Aeq1}) 
\begin{eqnarray}
\partial^{\lambda}A_{\lambda}(x)=0,  
\label{sup}
\end{eqnarray}
which guarantees that the independent freedom of vector fields 
is three. Now, inserting the relations among bosonic fields 
(\ref{SV})-(\ref{FA}) in (\ref{Bexp}) we obtain the bi-spinor 
solution
\begin{eqnarray}
\Psi(x)=\frac{1}{\sqrt{\mu}}
(\mu-\gamma^{\lambda}\partial_{\lambda})
(S(x)+\gamma_{5}\gamma^{\lambda}A_{\lambda}(x)). 
\label{Bsol}
\end{eqnarray}
Notice that this satisfies the BW equations (\ref{BWeq1})-(\ref{BWeq2}) 
automatically. 

Here we stress that the relations between fermionic solutions 
and bosonic solutions via bi-spinor field. The bi-spinor field 
is considered as bosonic fields in one side and as four spinor 
fields in the other side of (\ref{Bsol}):
\begin{eqnarray}
\Psi(x)
=(\psi^{(1)}(x),\psi^{(2)}(x),\psi^{(3)}(x),\psi^{(4)}(x)), 
\end{eqnarray}
where each spinor satisfies Dirac equation. Among four spinors 
$\psi^{(i)}(x)~(i=1,2,3,4)$, four components are independent 
which can be selected as
\begin{eqnarray}
\left(\begin{array}{cc}
\psi^{(1)}_{1} & \psi^{(2)}_{1}\\
\psi^{(1)}_{2} & \psi^{(2)}_{2}
\end{array}
\right), \label{four}
\end{eqnarray}
and other components are determined using BW equations. 
Explicitly, these four independent components in (\ref{four}) 
are written as $I$ and $\boldmath{\sigma}$, which correspond 
to anti-symmetric and symmetric parts in the original BW 
field ($\Psi^{(\rm BW)}=\Psi\gamma_{5}C$) as $\sigma_{y}$ 
and $\boldmath{\sigma}\sigma_{y}$. And these four components 
correspond to bosonic freedom: one for scalar $S$ and three 
for vectors $A_{k}$. Then the correspondence between fermions 
and bosons is uniquely established. 

Now, we write down the Lagrangian for the bi-spinor fields 
which gives us the BW equations (\ref{BWeq1})-(\ref{BWeq2}).
\begin{eqnarray}
\mathcal{L}
=-\frac{1}{8}{\rm Tr} \{ \bar{\Psi}(x)
(\gamma^{\lambda}\partial_{\lambda}+\mu)\Psi(x)
+{\Psi}(x)(\overleftarrow{\partial}_{\lambda}\gamma^{\lambda}
+\mu)\bar{\Psi}(x)\}, \label{L}
\end{eqnarray}
where the adjoint bi-spinor is defined as 
\begin{eqnarray}
\bar{\Psi}=(-i\gamma_{0}){\Psi}^{\dagger}(-i\gamma_{0}).
\end{eqnarray}
In terms of the bosonic fields $S(x)$ and $A_{\mu}(x)$, this 
Lagrangian is written as 
\begin{gather}
\mathcal{L}=-\mu^2S^{\dagger}(x)S(x) 
-\partial^{\lambda}S^{\dagger}(x)\partial_{\lambda}S(x)  \nn\\
-\mu^2A^{\lambda \dagger}(x)A_{\lambda}(x) 
-\frac{1}{2}F^{\lambda\tau \dagger}(x)F_{\lambda\tau}(x),
\end{gather}
which reproduces the correct equations of motion (\ref{Seq1})-(\ref{Aeq1})
for the bosonic fields.

The invariance of the Lagrangian (\ref{L}) under the phase 
transformation $\Psi(x)\rightarrow {\rm e}^{i\alpha(x)}\Psi(x)$ 
leads the conserved current in the bi-spinor expression 
\begin{eqnarray}
J_{\tau}=-\frac{\delta \mathcal{L}}{\delta\partial_{\tau}\alpha(x)}
=\frac{i}{4}{\rm Tr}\ \bar{\Psi}\gamma_{\tau}\Psi.
\label{currentBs}
\end{eqnarray}
In the bosonic expression we have 
\begin{eqnarray}
J_{\tau}=-i(S^{\dagger}\partial_{\tau}S-\partial_{\tau}S^{\dagger}S )
+i(A^{\lambda \dagger}F_{\lambda\tau}-F_{\lambda\tau}A^{\lambda}). 
\label{currentBo}
\end{eqnarray}
Clearly the current satisfies the conservation law 
\begin{eqnarray}
\partial^{\tau}J_{\tau}=0.
\end{eqnarray} 

\section{Superradiance puzzle between bosons and fermions 
in Kerr space-time}

We now apply BW formulation to the asymptotic flat space-time
in Kerr black hole and try to solve the superradiance puzzle 
between bosons and fermions. Consider the case for scalar 
bosons and the corresponding bi-spinors in flat space-time. 
We first express the scalar boson solution of Klein-Gordon 
equation (\ref{Seq1}) in the polar coordinate as 
\begin{eqnarray}
S(x)=Y^{m}_{\ell}(\theta, \phi)R_{\ell}(r)\exp{(-i\omega t)}, 
\label{Sspherical}
\end{eqnarray}
where $Y^{m}_{\ell}\, ,R_{\ell}$ and $\omega$ denote 
the spherical harmonics, the radial wave function and 
the frequency respectively. The corresponding bi-spinor 
solution is obtained using (\ref{Bsol}) and (\ref{Sspherical}),  
\begin{gather}
\Psi^{(\rm scalar)}(x)
=\frac{1}{\sqrt{\mu}}
(\mu-\gamma^{\lambda}\partial_{\lambda})S(x) \nn\\
=\frac{1}{\sqrt{\mu}}
\left(
\begin{array}{cc}
\mu+\omega, &~~i\boldmath{\sigma}\cdot \nabla \\
-i\boldmath{\sigma}\cdot \nabla, &~~\mu-\omega 
\end{array}
\right) Y^{m}_{\ell}(\theta,\phi)R_{\ell}(r)\exp{(-i\omega t)}. 
\label{p0}
\end{gather}
We write this as a set of four spinors,  
\begin{eqnarray}
\Psi^{(\rm scalar)}(x)
=\sqrt{\frac{\mu+\omega}{\mu}}(\psi^{(\omega \uparrow)}(x),
\psi^{(\omega \downarrow)}(x),
\psi^{(-\omega \uparrow)}(x),\psi^{(-\omega \downarrow)}(x)), 
\label{fourspinors}
\end{eqnarray}
where each suffix $(\omega \uparrow), \cdots (-\omega \downarrow)$ 
denotes the frequency and spin direction respectively. 

Among four spinors in (\ref{fourspinors}) we consider 
one solution $\Psi^{(\omega \uparrow)}$ because the 
net freedom is one \cite{note1}. The solution 
$\Psi^{(-\omega \uparrow)}$ is recombined in eigen-states 
of total angular momentum. For this purpose, the angular 
wave functions $Y^m_{\ell}$ can be written by 
the combination of normalized spin-angular functions 
$\mathcal{Y}$ as
\begin{eqnarray}
Y^{m}_{\ell}(\theta, \phi)
\left(
\begin{array}{c}
1\\0
\end{array}
\right)
=\sqrt{\frac{\ell+m+1}{2\ell+1}}\mathcal{Y}^{j_{3}}_{j,\ell}(\theta,\phi) 
-\sqrt{\frac{\ell-m}{2\ell+1}}\mathcal{Y}^{j_{3}}_{j',\ell}(\theta,\phi) , 
\end{eqnarray}
where the total angular momenta are $j=\ell+1/2$ and $j'=\ell-1/2$ 
with their azimuthal component $j_{3}=m+1/2$. Correspondingly 
the frequency $\omega$ with spin up spinor can be written 
as \cite{Saku67}
\begin{eqnarray}
\psi^{(\omega\uparrow)}=\sqrt{\frac{\ell+m+1}{2\ell+1}}\psi^{(\omega +)}(x)
-\sqrt{\frac{\ell-m}{2\ell+1}}\psi^{(\omega -)}(x)\ ,
\end{eqnarray}
where $\psi^{({\omega +})}$ and $\psi^{({\omega -})}$ denote 
the spin parallel and antiparallel spinors defined by  
\begin{gather}
\psi^{({\omega +})}=\frac{1}{r}
\left(
\begin{array}{c}
F(r)\mathcal{Y}^{j_{3}}_{j,\ell}\\
iG^{(+)}(r)\mathcal{Y}^{j_{3}}_{j,\ell+1} 
\end{array}
\right)\rm{e}^{-i\omega t},  \nn\\ 
\psi^{({\omega -})}=\frac{1}{r}
\left(
\begin{array}{c}
F(r)\mathcal{Y}^{j_{3}}_{j',\ell}\\
iG^{(-)}(r)\mathcal{Y}^{j_{3}}_{j',\ell-1} 
\end{array}
\right)\rm{e}^{-i\omega t},  \nn\\
F(r)=\sqrt{\mu+\omega}\, r\,R_{\ell}(r),  \nn\\ 
G^{(+)}(r)=\frac{(\partial_{r}-({\ell}+1)/{r})F(r)}{\omega+\mu},
~~~G^{(-)}(r)=\frac{(\partial_{r}+\ell/r)F(r)}{\omega+\mu}. 
\end{gather}
This establishes the relation between scalar wave function $S$ 
and spinor wave function $\psi^{(\omega\uparrow)}$ or $\psi^{(\pm)}$ 
using the bi-spinor field $\Psi$ explicitly. 

Next we consider the radial component of the conserved current 
for scalar boson \cite{note2},
\begin{eqnarray}
J^{(\rm{scalar})}_{r}
&=&\int r^2\sin{\theta}d\theta d\phi (-i)(S^{(0)\, *}\partial_{r}S^{(0)}
-\partial_{r}S^{(0)\, *}S^{(0)})/N_{\rm B} \nonumber\\
&=&r^2i(\partial_{r} R^{*}_{\ell}R_{\ell}-R^{*}_{\ell}
\partial_{r}R_{\ell}). 
\label{Bcurrent}
\end{eqnarray}
For spinors we have \cite{note3}
\begin{eqnarray}
J^{(k)}_{r}
=\int r^2\sin{\theta}d\theta d\phi 
\ i\bar{\psi}^{(k)}\, {\boldmath{\gamma}}\cdot
\frac{{\boldmath{x}}}{r}\, \psi^{(k)}/N_{\rm F}
=r^2 i(\partial_{r} R^{*}_{\ell}R_{\ell}-R^{*}_{\ell}\partial_{r}R_{\ell}), 
\label{Fcurrent} 
\end{eqnarray}
where suffix $k$ denotes $(\omega \uparrow), (\omega +)$ or 
$(\omega -)$ respectively. Then we obtain the current relation 
between scalar boson and spinor (spin up, spin parallel 
or spin anti-parallel) as 
\begin{eqnarray}
J^{(\rm{scalar})}_{r}
=J^{(\omega \uparrow)}_{r}=J^{(\omega +)}_{r}=J^{(\omega -)}_{r}.
\label{BFcurrent}
\end{eqnarray}
Note that the expression for each current does not depend on 
the wave function normalization factors (see footnotes $b$ 
and $c$).

In order to solve the superradiance puzzle applying the BW 
formulation, we study the scattering problems for fermionic 
and bosonic fields in Kerr black hole space-time in 
Boyer-Lindquist coordinates \cite{Boyer1967} 
\begin{gather}
ds^2=\frac{\Delta}{\Sigma}[dt-a\sin^{2}{\theta}d\phi]^2
+\frac{\Sigma}{\Delta}dr^2+\Sigma d\theta^2
+\frac{\sin^2{\theta}}{\Sigma}[(r^2+a^2)d\phi-adt]^2, \nn\\
\Delta=r^2-2Mr+a^2,~~~\Sigma=r^2+a^2\cos^2{\theta},
\end{gather}
where $M$ and $a$ denote the mass and angular momentum of the Kerr 
black hole respectively. For the scattering problem for the spin 0 
scalar filed, we let
\begin{eqnarray}
S(x)=R(r)Y(\theta, \phi)~{\exp}{(-i\omega t)},
\end{eqnarray}
where $R(r)$ and $Y(\theta, \phi)$ denote the radial and  
angular wave functions respectively. They obey the field 
equations
\bea
\Big(\dfrac{1}{\sin{\theta}} 
\partial_{\theta}\sin{\theta}\partial_{\theta}
-(a\omega\sin{\theta}-\dfrac{m}{\sin{\theta}})^2-\mu^2a^2\cos^2{\theta} 
+\nu \Big)Y(\theta, \phi)&=&0, \\
\Big(\partial_{r}\Delta\partial_{r}
+\dfrac{\big((r^2+a^2)\omega-am \big)^2}{\Delta}
-\mu^2r^2 -\nu \Big)R(r)&=&0, 
\eea
where $\mu$ and $\nu$ denote the mass of particle and 
separation parameter. 

To study the behavior of radial wave function near infinity 
and event horizon we introduce a new radial coordinate $r_*$, 
\begin{eqnarray}
\frac{dr_*}{dr}=\frac{r^2+a^2}{r^2-2M r+a^2}.
\end{eqnarray}
Using the new coordinate, the radial field solutions in Kerr 
space-time become free waves near the spatial infinity 
$r\rightarrow \infty~(r_{*}\rightarrow {\infty})$  
and event horizon $r\rightarrow r_{\rm H}
=M+\sqrt{M^2-a^2}~(r_{*}\rightarrow {-\infty})$,
\begin{eqnarray}
\sqrt{r^2+a^2}\,R(r) \sim 
\left\{
\begin{array}{cc}
\dfrac{A^{(\rm inc)}_{\rm B}\exp{(-ip_{\infty}r_*)}
+A^{(\rm ref)}_{\rm B}\exp{(ip_{\infty}r_*)}}{\sqrt \omega}
~~~~~(r\rightarrow \infty),\\
\dfrac{A^{(\rm trans)}_{\rm B}\exp{(-ip_{\rm H}r_*)}}
{\sqrt {\mid p_{\rm H}\mid}}~~~~~(r\rightarrow r_{\rm H}),
\end{array}
\right. 
\label{Sasymptoticsol}
\end{eqnarray}
where $A^{(\rm inc)}_{\rm B}$, $A^{(\rm ref)}_{\rm B}$, 
and $A^{(\rm trans)}_{\rm B}$ denote the incident, 
reflected, and transmitted waves respectively. Momenta 
near the infinity and event horizon are denoted by 
\begin{eqnarray}
p_{\infty}=\sqrt{\omega^2-\mu^2}\ \ {\mbox{and }} \ \ 
p_{\rm H}=\omega-m \Omega_{\rm H},
\end{eqnarray}
where particle mass $\mu$, magnetic quantum momentum $m$ 
and angular velocity $\Omega_{\rm H}=a/(r_{\rm H}^2+a^2)$ 
respectively. Note that momentum $p_{\rm H}$ is equivalent 
to the effective energy because particles are treated as 
massless near horizon. Using the asymptotic solution for 
scalar boson (\ref{Sasymptoticsol}), the current relation 
of the radial component can be derived for the infinity 
and event horizon region,
\begin{eqnarray}
\frac{p_{\infty}}{\omega}
(\mid A^{(\rm inc)}_{\rm B}\mid^2-\mid A^{(\rm ref)}_{\rm B}\mid^2)
= \frac{p_{\rm H}}{\mid p_{\rm H}\mid} 
\mid A^{(\rm trans)}_{\rm B}\mid^2. 
\label{CCR1}
\end{eqnarray}
The momentum at infinity $p_{\infty}$ is positive definite 
by definition of incident and reflected waves. From this we 
have the superradiance condition for scalar bosons 
\begin{eqnarray}
\omega \ p_{\rm H}<0. 
\label{srcondition}
\end{eqnarray} 
This superradiance condition taking account only of scalar 
bosons is consistent with previous works on the second 
quantized method by Unruh \cite{Unruh74}, general analytical 
method using the Teukolsky equation in the massless limit 
by Mano and Takasugi \cite{Taka97}, and Hawking's thermodynamic 
theorem of black hole area by Bekenstein \cite{Beken73,note4}. 
The condition (\ref{srcondition}) is realized in the cases 
(type 1) $\omega>0$ and $p_{\rm H}<0$ or (type 2) $\omega<0$ 
and $p_{\rm H}>0$.

Next we consider the scattering problem for the spin 1/2 spinor 
field in Kerr space-time. Write the spinor wave function in the 
form,
\begin{eqnarray}
\psi(x)=\frac{1}{(\Delta\Sigma)^{1/4}}
\left(
\begin{array}{c}
F(r)\mathcal{Y}(\theta,\phi)\\
iG(r)\mathcal{Y}'(\theta,\phi)
\end{array}
\right),
\end{eqnarray}
where $\mathcal{Y}(\theta,\phi),~\mathcal{Y}'(\theta,\phi)$ 
stand for normalized spin-angular functions. Asymptotic 
solutions of radial components are obtained 
\begin{eqnarray}
F(r) &\sim& 
\sqrt{\frac{\omega+\mu}{\omega}}(
A^{(\rm inc)}_{\rm F}\exp{(-ip_{\infty}r_*)}
+A^{(\rm ref)}_{\rm F}\exp{(ip_{\infty}r_*)}), \\
G(r) &\sim& \frac{1}{(\omega+\mu)}\frac{d}{dr_*}F(r), 
\end{eqnarray}
for infinity  $(r\rightarrow \infty)$ and 
\begin{eqnarray}
F(r) &\sim& 
A^{(\rm trans)}_{\rm F}\exp{(-ip_{\rm H}r_*)}, 
\label{Fsolevent1}\\
G(r) &\sim& iF(r), 
\label{Fsolevent2}
\end{eqnarray}
for near event horizon $(r\rightarrow r_{\rm H})$, 
where $A^{(\rm inc)}_{\rm F}$, $A^{(\rm ref)}_{\rm F}$, 
and $A^{(\rm trans)}_{\rm F}$ denote fermionic amplitudes 
of incident wave, reflected, and transmitted waves 
respectively. The radial current relation for spinors 
between the infinity and event horizon is given by
\begin{eqnarray}
\frac{p_{\infty}}{\omega} (\mid A^{(\rm inc)}_{\rm F}\mid^2
-\mid A^{(\rm ref)}_{\rm F}\mid^2)
= \mid A^{(\rm trans)}_{F}\mid^2.  
\label{CCR2}
\end{eqnarray}
Note that the normalization factors of spinor solution 
in (\ref{Fsolevent1}) for near horizon is one, and as a 
result the factor in front of $\mid A^{(\rm trans)}_{\rm F}\mid$ 
is also one (the speed of light), which is consistent 
with the massless neutrino theory. From this we conclude 
that the superradiance cannot occur for $0<\omega$ and 
$p_{\rm H}<0$ (type 1). The result is completely consistent 
with the previous works on the second quantization method 
with respect to the exclusion principle which states that 
all states with $p_{\rm H}<0$ are filled by Unruh and 
others \cite{Unruh74,Maeda76,Wagh85}. 

In order to relate the bosonic and fermionic current, 
we apply the BW formulation for spacial infinity, which 
can be treated as flat space-time. Using (\ref{BFcurrent})
we obtain the relation at spacial infinity $(r\rightarrow \infty)$,  
\begin{eqnarray}
\mid A^{(\rm inc)}_{\rm B}\mid^2-\mid A^{(\rm ref)}_{\rm B}\mid^2
=\mid A^{(\rm inc)}_{\rm F}\mid^2-\mid A^{(\rm ref)}_{\rm F}\mid^2. 
\label{CCR3}
\end{eqnarray}
Combining three conserved current relations (\ref{CCR1}), 
(\ref{CCR2}), and (\ref{CCR3}), we obtain the current relation 
near the horizon as
\begin{eqnarray}
\frac{p_{\rm H}}{\mid p_{\rm H}\mid} 
\mid A^{(\rm trans)}_{\rm B}\mid^2
=\mid A^{(\rm trans)}_{\rm F}\mid^2, 
\end{eqnarray}
which shows that the momentum near the event horizon is 
positive,  
\begin{eqnarray}
0< p_{\rm H}=\omega-m \Omega_{\rm H}.
\end{eqnarray} 
This means that the superradiance of $0< \omega$ and 
$p_{\rm H}<0$ (type 1) can not occur for both bosons 
and fermions. It should be noted that the superradiance 
of $\omega<0$ and $p_{\rm H}>0$ (type 2) is consistent 
with all current relations of (39), (46), and (47). 

\section{Summary}

We have studied the superradiance puzzle between bosons and 
fermions in Kerr black hole space-time using the Bargmann-Wigner 
formulation, and established the direct wave function relation 
between massive scalar bosons and spinors via bi-spinor fields 
With this we have demonstrated that, just as the fermionic case,  
the superradiance for scalar bosons of the type $0<\omega$ 
and $p_{\rm H}=\omega-m\Omega_{H}<0$ does not occur in Kerr 
space-time. This result is consistent with that in (2+1)-dimensional 
analysis \cite{Kenmoku08, Kenmoku08-2}.  

However, it should be emphasized that the superradiance of 
$\omega <0$ and $0< p_{\rm H}$ (type 2) is still possible 
(the magnetic quantum number $m$ should be negative in this 
case) \cite{note5}. In type 2 case, the negative energy particle 
goes into black hole with positive momentum near horizon 
($p_{\rm H}>0$), which corresponds to the interpretation to 
get energy from black hole and can occur the superradiance 
phenomena. 

In case of $\omega<0 $ and $0<p_{\rm H}$, we use 
$\psi^{(-\omega\downarrow)}$ instead of $\psi^{(\omega\uparrow)}$ 
as an independent spinor among four spinors in (\ref{fourspinors}) 
and even in this case the current relations (\ref{CCR1}), 
(\ref{CCR2}), and (\ref{CCR3}) hold in the same way. These 
current relations are consistent with the superradiance of 
(type 2) $\omega<0$ and $0< P_{\rm H} $. In this case the 
incoming particles with negative frequency get enough energy 
from black hole and are scattered backward strongly when they 
have opposite angular momentum to that of black hole.

This is very important because in quantum field theory
the negative frequency component becomes an essential part 
of the field. Moreover, the physical vacuum is made of infinite 
number of virtual particle-antiparticle pairs, so that the 
incoming particle is accompanied by a large number of virtual 
particle-antiparticle pairs. So near the horizon, the 
virtual antiparticles (the negative frequency component) 
can be absorbed by the blackholes. Physically this could make 
the reflected flux larger than the incident flux at the horizon, 
resulting in the superradiance \cite{Pen69}. 

This naturally relates our result to the Hawking 
radiation \cite{Hawking74,Hawking75}. Of course the Hawking 
radiation is a quantum effect, and the superradiance is 
a classical phenomenon. Nevertheless the above discussion 
tells that our result is not inconsistent with the Hawking 
radiation. It suggests that the Hawking radiation could 
be interpreted as a superradiadion of the virtual 
particle-antiparticle pairs.

We can extend our method for the superradiance problem of 
vector bosons using similar analysis to scalar bosons 
which will be done in a separate paper.

\section*{Acknowledgements}

MK thanks Professor Kazuyasu Shigemoto for usefull 
discussions. YMC is supported in part by the Basic 
Science Research Program through the National Research 
Foundation (Grant 2012-002-134) of the Ministry of Science 
and Technology and by Konkuk University.



\begin{thebibliography}{99}
\bibitem{note1} The current is normalized by the total scalar density,
$N_{\rm B}=\int d^3x (-i)
(S^{(0)*}\partial^{t}S^{(0)}-\partial^{t}S^{(0)*}S^{(0)})$.
By this expression of current, the wave function normalization 
becomes irrelevant.
\bibitem{note2} The current is normalized by the total scalar density,
$N_{\rm B}=\int d^3x (-i)(S^{(0)*}\partial^{t}S^{(0)}
-\partial^{t}S^{(0)*}S^{(0)})$. By this expression of current, 
the wave function normalization becomes irrelevant.
\bibitem{note3} The current is normalized by the total spinor density, 
$N_{\rm F}=\int d^3x i\bar{\psi}^{(k)}\gamma^{0}\psi^{(k)}$.
\bibitem{note4} Bekenstein derives the relation 
$dM (1-m\Omega_{\rm H}/\omega)>0$ from the Hawking's 
thermodynamical theorem with the conservation law of 
energy and angular momentum, and derives the superradiance 
condition for $dM<0$ as $1-m\Omega_{\rm H}/\omega<0$. This 
is coincident with our result (\ref{srcondition}) for both 
positive and negative $\omega$.
\bibitem{note5} The theory is invariant under the reflection 
symmetry with respect to frequency and azimuthal number 
$(\omega, m)\leftrightarrow(-\omega, -m)$; one of 
which is physical ($\omega <0, p_{\rm H}>0$: type 2)
and the other is unphysical ($\omega >0, p_{\rm H}<0$: type 1).
\end{thebibliography}
\end{document}